\documentclass[twocolumn,floatfix,superscriptaddress,jcp]{revtex4-1}

\usepackage[utf8]{inputenc}
\usepackage{natbib}
\usepackage{graphicx}
\usepackage{xcolor}
\usepackage{bm}
\usepackage{mhchem}
\usepackage{physics}
\usepackage{amsfonts}
\usepackage[normalem]{ulem}
\usepackage{comment}
\usepackage[normalem]{ulem}
\usepackage{siunitx}
\usepackage{pdfpages}
\usepackage{pgffor}
\makeatletter
\AtBeginDocument{\let\LS@rot\@undefined}
\makeatother

\definecolor{cocol}{rgb}{0,0.6,0}

\definecolor{bluish}{rgb}{0,0.4,0.9}
\newcommand{\rev}[1]{#1}
\newcommand{\revrev}[1]{#1}

\makeatletter
\g@addto@macro\normalsize{%
}
\makeatother

\usepackage{physics}
\usepackage{xparse}

\newcommand{\mbf}[1]{\ensuremath{\mathbf{#1}}}

\NewDocumentCommand{\rep}{s d<| d|>}{%
\IfBooleanTF{#1}{
   \IfValueTF{#2}{
       \IfValueTF{#3}{\braket{#2}{#3}}{\bra{#2}}
       }{
       \IfValueTF{#3}{\ket{#3}}{}
       }
   }{
   \IfValueTF{#2}{
       \IfValueTF{#3}{\braket*{#2}{#3}}{\bra*{#2}}
       }{
       \IfValueTF{#3}{\ket*{#3}}{}
       }
   }
}

\NewDocumentCommand{\rbra}{sm}{\IfBooleanTF{#1}{\rep*<#2|}{\rep<#2|}}
\NewDocumentCommand{\rket}{sm}{\IfBooleanTF{#1}{\rep*|#2>}{\rep|#2>}}
\NewDocumentCommand{\rbraket}{smom}{
    \IfBooleanTF{#1}{
        \IfNoValueTF{#3}{\rep*<#2||#4>}{\rep*<#2|#3\rep*|#4>}
    }{
        \IfNoValueTF{#3}{\rep<#2||#4>}{\rep<#2|#3\rep|#4>}
    }
}

\NewDocumentCommand{\field}{o m e{_} e{^} o e{_} e{^}}{
\IfValueTF{#5}{\overline{
  #2\IfValueT{#3}{_#3}\IfValueT{#4}{^{\otimes #4}} %
  \otimes
  #5\IfValueT{#6}{_#6}\IfValueT{#7}{^{\otimes #7}} %
  \IfValueT{#1}{;#1}
}}{
  \IfValueTF{#4}{\overline{
     #2\IfValueT{#3}{_#3}\IfValueT{#4}{^{\otimes #4}}
     \IfValueT{#1}{;#1}
  }}
  {#2\IfValueT{#3}{_#3}}
}
}

\NewDocumentCommand{\frho}{o e{_} e{^}}{
\field[#1]{\rho}_{#2}^{#3}
}

\newcommand{\e}{a}  %

\newcommand{\bx}{\mbf{x}}

\NewDocumentCommand{\ex}{e_}{
\IfValueTF{#1}{\e_{#1}\bx_{#1}}{\e\bx}
}  %

\NewDocumentCommand{\lm}{e_}{
\IfValueTF{#1}{l_{#1}m_{#1}}{lm}
}
\NewDocumentCommand{\nlm}{e_}{
\IfValueTF{#1}{n_{#1}\lm_{#1}}{n\lm}
}
\NewDocumentCommand{\enlm}{e_}{
\IfValueTF{#1}{\e_{#1}\nlm_{#1}}{\e\nlm}
}
\NewDocumentCommand{\en}{e_}{
\IfValueTF{#1}{\e_{#1}n_{#1}}{\e n}
}
\NewDocumentCommand{\nlk}{e_}{
\IfValueTF{#1}{n_{#1}l_{#1}k_{#1}}{nlk}
}
\NewDocumentCommand{\enlk}{e_}{
\IfValueTF{#1}{\e_{#1}\nlk_{#1}}{\e\nlk}
}
\NewDocumentCommand{\enl}{e_}{
\IfValueTF{#1}{\en_{#1}l_#1}{\en l}
}

\NewDocumentCommand{\nnl}{s}{
\IfBooleanTF{#1}{n_1 n_2 l}{n_1; n_2; l}
}
\NewDocumentCommand{\ennl}{s}{
\IfBooleanTF{#1}{\en_1 \en_2 l}{\en_1; \en_2; l}
}

\NewDocumentCommand{\gslm}{s}{
\IfBooleanTF{#1}{\sigma\lambda\mu}{\sigma;\lambda\mu}
}

\newcommand{\nmax}{n_\text{max}}
\newcommand{\lmax}{l_\text{max}}

\begin{document}

\title{Comment on ``Manifolds of quasi-constant SOAP and ACSF fingerprints and the resulting failure to
machine learn four body interactions"}

\author{Sergey N. Pozdnyakov}
\affiliation{Laboratory of Computational Science and Modelling, Institute of Materials, Ecole Polytechnique F\'ed\'erale de Lausanne, Lausanne 1015, Switzerland}
\author{Michael J.~Willatt}
\affiliation{Laboratory of Computational Science and Modelling, Institute of Materials, Ecole Polytechnique F\'ed\'erale de Lausanne, Lausanne 1015, Switzerland}
\author{Albert P. Bart\'ok}
\affiliation{Department of Physics and Warwick Centre for Predictive Modelling, School of Engineering, University of Warwick, Coventry CV4 7AL, United Kingdom}
\author{Christoph Ortner}
\affiliation{Mathematics Institute, University of Warwick, Coventry CV4 7AL, United Kingdom}
\author{G\'abor Cs\'anyi}
\affiliation{Engineering Laboratory, University of Cambridge, Trumpington Street, Cambridge CB2 1PZ, United Kingdom}
\author{Michele Ceriotti}
\email{michele.ceriotti@epfl.ch}
 \affiliation{Laboratory of Computational Science and Modelling, Institute of Materials, Ecole Polytechnique F\'ed\'erale de Lausanne, Lausanne 1015, Switzerland}
\date{\today}

\begin{abstract}
The ``quasi-constant'' SOAP and ACSF fingerprint manifolds recently \rev{discovered} by Parsaeifard and Goedecker\cite{pars-goed22jcp} are a direct consequence of the presence of degenerate pairs of configurations, a known shortcoming of all low-body-order atom-density correlation representations of molecular structures. 
Contrary to the configurations that are rigorously singular \revrev{-- that we demonstrate can only occur in finite, discrete sets --} the continuous ``quasi-constant'' manifolds exhibit low, but non-zero, sensitivity to atomic displacements. 
Thus, it is possible to build \revrev{interpolative} machine-learning models of high-order interactions along the  manifold, \revrev{even though the numerical instabilities associated with proximity to the exact singularities affect the accuracy and transferability of such models, to an extent that depends on numerical details of the implementation. }
\end{abstract}

\maketitle

The paper by Parsaeifard and Goedecker\cite{pars-goed22jcp} presents an interesting observation on the numerical behaviour of smooth overlap of atomic positions (SOAP) and atom-centered symmetry functions (ACSF) representations. Manifolds along which the representation exhibits low sensitivity to atomic displacements (identified by small singular values of the Jacobian $\mathbf{J}$) are described, and it is shown that the performance of machine-learning (ML) models based on these features is poor when the target has a  four-body character. Representations which are inherently high-order, such as overlap matrix fingerprints, do not exhibit these numerical issues. 
The significance of these empirical observations cannot be appreciated without putting them in the context of the fundamental properties of atom-correlation based representations. 
Here we show that: (1) Low-sensitivity regions are a direct manifestation of the degenerate manifolds discussed in Ref.~\citenum{pozd+20prl}, which lead to configurations where the Jacobian is \emph{exactly} singular\cite{pozd+21ore}; (2) The extent to which these singular configurations induce a region of low sensitivity depends on the numerical details of the implementation; (3) Although low sensitivity affects the extrapolative power of the model, it does not altogether prevent model fitting \revrev{in the interpolative regime}. %

\begin{figure}[tb]
    \centering
    \includegraphics[width=0.85\linewidth]{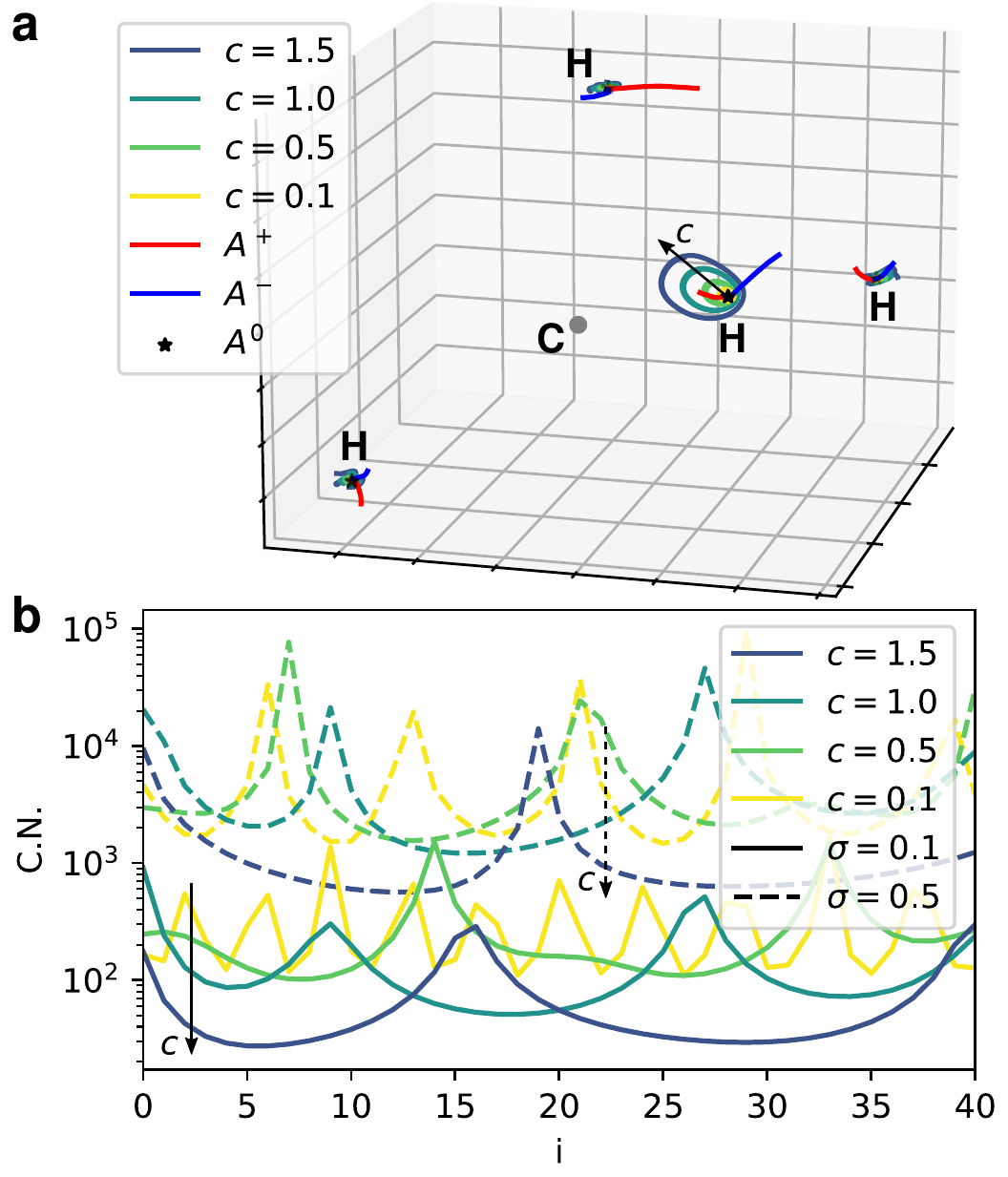}
\caption{(a) Low-sensitivity manifolds computed starting at different distances from a first-order degenerate \ce{CH4} environment, $A^0$. 
The discrete degenerate pair manifolds $A^\pm$ are also shown. The manifold with $c=1$ \rev{is graphically indistinguishable from the manifold used by Parsaeifard and Goedecker, although it depends slightly on the SOAP implementation and parameters.}
The others start from a point with a distance to $A^0$ scaled by the given factor $c$. 
(b) Condition number (CN) of $\mathbf{J}$ for structures along the low-sensitivity manifolds. Full (dashed) lines correspond to a smearing $\sigma=0.1$ (0.5)~\AA{} smearing. Even though the CN along the manifold varies depending on $\sigma$, the geometries are very similar.
SOAP features are computed using librascal\cite{musi+21jcp}, with $\nmax=\lmax=8$, using optimal radial basis functions\cite{gosc+21jcp}. }
    \label{fig:manifolds}
\end{figure}

\begin{figure}[b]
    \centering
    \includegraphics[width=1.0\linewidth]{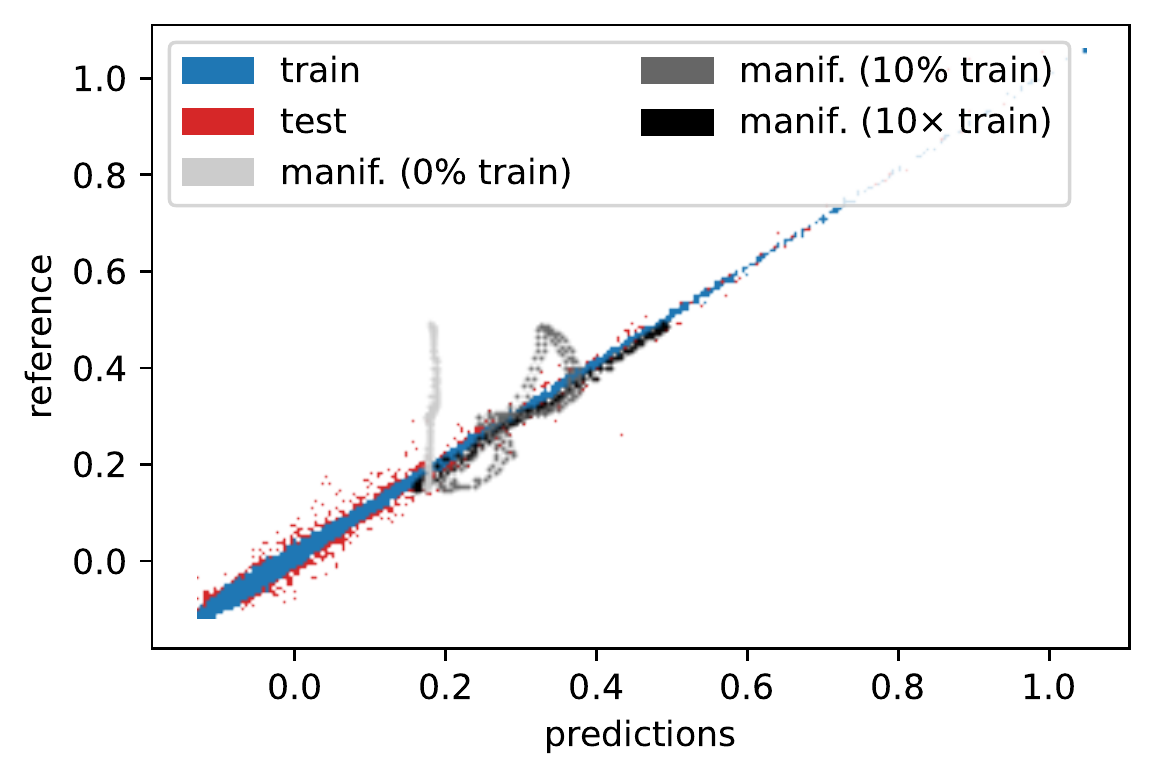}
    \caption{Parity plot for neural-network models based on SOAP features, for \revrev{a manifold built following the protocol from Parsaeifard and Goedecker, with SOAP features having $\sigma=0.2$~\AA{}, $\lmax=8$ angular channels and  $\nmax=8$  optimal radial basis functions}. Blue and red points correspond to training and validation sets. The light gray points are the predictions of for the quasi-degenerate manifold of a model trained without including any manifold configuration. The darker gray points are the predictions for a model that includes 20 (10\%) of the manifold configurations in the training set. The black points are the predictions for a model that includes the full manifold during training, repeated \revrev{10 times to trigger an overfitting behavior.} Changes for the validation error between models are small. }
    \label{fig:nn-experiments}
\end{figure}

Given that ML interatomic potentials for molecules and especially materials are usually built as a sum of atomic contributions~\cite{behl21cr}, the geometry of atom-centered environments are described by symmetrized representations. Many of these can be understood in terms of correlations of the neighbor density \cite{will+19jcp}, and classified depending on the order of correlations $\nu$ that they incorporate: representations that have the same $\nu$ may differ in computational efficiency and numerical stability, but encode the same type of information. 
Recently, a fundamental limitation of representations that rely on 3-body correlations (such as SOAP\cite{bart+13prb} and ACSF\cite{behl-parr07prl,behl11jcp}) was revealed.\cite{pozd+20prl}
Pairs of environments exist  which share \emph{identical} 3-body representations even though the configurations differ.
The existence of indistinguishable configurations affects the expressive power of models based on those features, which will be incapable of predicting distinct values for the corresponding atom-centered properties, \revrev{\emph{even if both structures are used during training.}}

\rev{One can understand the origin of these degeneracies by noting that, in the complete basis set limit, $\nu=2$ features contain information on the distances between the central atom and two of its neighbors, as well as the angle between them. 
Thus, they contain enough information to compute the set of entries of the Gram matrix for the environment. 
Much as it is the case for distance-based matrix representations of molecules\cite{rupp+12prl}, the Gram matrix allows to reconstruct a point cloud. 
Crucially, however, $\nu=2$ features are invariant to atom index permutations, and do not allow to recover the position of the entries in the matrix, similar to what is seen when symmetrizing distance-matrix representations by sorting their entries\cite{mous12prl}.
Most of the time, shuffling the entries of the Gram matrix leads to point clouds that can only be embedded in a dimension higher than 3. 
Degenerate structures occur when two or more permutations result in distinct 3D point clouds.
Incidentally, this discussion proves that a continuous manifold along which $\nu=2$ features are \emph{exactly} constant cannot exist: for a finite number of neighbors there can be only a finite number of permutations of the entries in the Gram matrix of an environment and hence a finite set of degenerate structures sharing the same $\nu=2$ features. 
The largest known set of discrete degeneracies contains three structures.\cite{pozd+20prl} 
}

Getting back to the discrete degenerate pairs of Ref.~\citenum{pozd+20prl}, one can find that matching pairs are arranged along two continuous manifolds, $A^\pm$: at their intersection one finds structures $A^0$ for which $\mathbf{J}$ has a ``spurious'' singular value -- the features change as the square of the atomic displacement away from $A^0$, even if $A^0$ is not symmetric.\cite{pozd+21ore} 
Both ``zero-order'' degenerate pairs $A^\pm$ and ``first-order'' singular configurations $A^0$ are intrinsic problems of any 3-body representation.
The ``quasi-constant'' manifolds observed by Parsaeifard and Goedecker are not a distinct, unrelated pathology, but a direct consequence of \rev{the structures being geometrically close} to a first-order singularity, the spurious zero-sensitivity structure at the intersection of the manifolds of degenerate pairs.
As shown in Fig.~\ref{fig:manifolds}a, one can find a pair of discrete degeneracies that meet close to the centre of the low-sensitivity manifold. \rev{Given that there is an entire continuous manifold of first-order singularities, one can expect to find low-sensitivity regions for several other molecular geometries.}
For each $A^0$, many ``low-sensitivity orbits'' can be found by following the procedure outlined in Ref.~\citenum{pars-goed22jcp} starting from points in the vicinity of the singularity.
\revrev{ The level of sensitivity depends on the exact starting point, with the condition number of $\mathbf{J}$ usually increasing as one approaches $A^0$.}
This observation also explains the strong dependency of the sensitivity on the parameters of the representation, which is noted in Ref.~\citenum{pars-goed22jcp}. 
The use of large Gaussian smearing $\sigma$, or a small basis set that is not capable of discretizing accurately the neighbor-density correlations, means that the presence of a first-order singularity affects a larger region of configurations, with high condition numbers observed also further from $A^0$ (Fig.~\ref{fig:manifolds}b). 

Since the ``quasi-constant'' manifold is close, but does not intersect, the discrete degeneracies, it should be possible to distinguish structures along it, and train a model for the four-body potential, provided that one uses a high-resolution representation, and includes structures close to the manifold in the training set. 
\revrev{To verify this, we perform similar numerical experiments to those discussed in Ref.~\citenum{pars-goed22jcp},  using an orbit starting at the same distance from $A_0$ as for the manifold discussed by Parsaeifard and Goedecker, following the lowest-sensitivity direction for SOAP features with $\nmax=8$ optimal radial functions,\cite{gosc+21jcp} $\lmax=8$, $\sigma=\SI{0.2}{\angstrom}$, which leads to a condition number comparable to the ``tight'' settings shown in Fig. 5 of Ref.~\citenum{pars-goed22jcp}.} 
We use a simple dense neural network with two inner layers with $80$ neurons in each, \rev{similar to that discussed in Ref.~\citenum{pars-goed22jcp}.}
Hyperbolic tangents are used as activation, and group norms\cite{wu2018arxiv} inserted to accelerate convergence (Fig.~\ref{fig:nn-experiments}). 
SOAP features \emph{do} perform rather poorly in the extrapolative regime, which is unsurprising given that the ``quasi-constant'' manifold is in the close vicinity of the intersection of the manifolds of degenerate pairs, far from the training set, and that the target is specifically built as a pure 4-body term, which is intrinsically challenging for a 3-body representation. 
However, contrary to what is claimed in~Ref.~\citenum{pars-goed22jcp}, a SOAP-based model \emph{can} approximate the energy along the ``quasi-constant'' manifolds, if a few reference configurations from it are included as part of the training set (Fig.~\ref{fig:nn-experiments}, dark gray points). \revrev{If the full manifold is included in the training set, increasing its weight to trigger an over-fitting behavior, (black points), the manifold properties can be estimated almost perfectly. Manifolds that are even closer to the singularity, or that are built with higher Gaussian smearing, require even denser sampling and/or more flexible models.  Even though the purely interpolative regime makes this an artificial exercise,  which does not address the underlying problems of low-order atom-centered representations, it does prove that the quasi-constant nature of the manifold does not prevent learning {\em in principle}, contrary to the case of exact degeneracies.}

This is not to say that the \rev{numerical experiments in Ref.~\citenum{pars-goed22jcp} are} without merit: numerical stability and extrapolative power are important, although different from the outright failure of constructing a regression model.
A meaningful analysis of regression performance would require working with realistic materials data, for which a careful tuning of the sensitivity of the representation often improves the accuracy of models\cite{huan-vonl16jcp,will+18pccp,caro19prb}. 
Furthermore, even though 3-body features \emph{do} exhibit very severe pathological behaviour in the form of degenerate pairs of configurations, this does not necessarily lead to catastrophic failures of the corresponding ML potentials. The saving grace, that appears to underpin the enormous success of \rev{ML models based on $\nu=1,2$ features,} is the fact that most models are built using \emph{multiple atoms as centers}, which resolve the degenerate pairs discussed in Ref.~\citenum{pozd+20prl} -- and thus largely address the underlying cause of the low sensitivity observed by Parsaeifard and Goedecker.~\cite{pars-goed22jcp}
\revrev{The benchmarks in Ref.~\citenum{pozd+20prl} provide an estimate of the combined impact of \emph{all types of degeneracies} for a challenging dataset of 7 million \ce{CH4} configurations, whose atomization energies span a range of several tens of eV. When using a multi-center model, the limiting accuracy that can be reached by a NN model based on SOAP features is of the order of 1~kcal/mol.  }

\rev{
The problem of constructing a complete and concise set of descriptors for atomic configurations is not entirely solved yet. Low-body order features are known to present pathological behavior for some structures, and can be systematically improved by incorporating higher-order terms. Some schemes, such as MTP\cite{shap16mms}, ACE\cite{drau19prb}, NICE\cite{niga+20jcp} give in principle access to a \emph{linearly complete} representation. Others, such as OM fingerprints\cite{sade+13jcp} do not generate a complete linear basis,\cite{musi+21cr} but can incorporate enough higher-order terms to resolve all known degeneracies (see Section 5 of Ref.~\citenum{musi+21cr} for a more thorough discussion). }

\revrev{
It is important to distinguish fundamental shortcomings connected with the geometric nature of the  density correlations from the numerical instabilities they cause for neighboring configurations.
For the former, SOAP, ACSF, or any $\nu=2$ construction such as \rev{DeepMD\cite{zhan+18prl} or the 3-body FCHL}\cite{fabe+18jcp} would fail in exactly the same way regardless of the hyperparameters and the implementation details, and lead to the impossibility of learning properties even for structures included in the train set. 
For the latter, tuning the hyperparameters of the representation, choosing a well-converged basis set and a flexible regression model can mitigate the impact of the instabilities, even though impractical amounts of training data and unusual training protocols might be needed to achieve accurate predictions.}
Incorporating higher-body order terms, either with systematically-improvable density-correlation schemes\cite{shap16mms,drau19prb,niga+20jcp} or with OM fingerprints\cite{sade+13jcp} should be attempted any time one suspects that these effects are impacting the performance of a ML model in a practical application.

\section*{Supporting materials}

A docker file that uses SOAP features computed from librascal\cite{LIBRASCAL,musi+21jcp} to reconstruct the quasi-constant orbits, and train a simple NN based on PyTorch\cite{pytorch} is available from \url{https://github.com/lab-cosmo/ch4-manifold-training}.

\begin{acknowledgments}
MJW, SP and MC acknowledge funding by the Swiss National Science Foundation (Project No. 200021-182057) and by the Swiss Platform for Advanced Scientific Computing (PASC). CO acknowledges support by the Natural Sciences and Engineering Research Council of Canada (NSERC) [funding reference number GR019381].
\end{acknowledgments}

\end{document}